# Magnetization study to probe the effect of $N_2^+$ irradiation on Fe/Co bilayers

Harkirat Singh[1], Ratnesh Gupta[2], Ashish Khandelwal[3] and Chiranjib Mitra[1*]

[1]Indian Institute of Science Education and Research, Kolkata, Mohanpur Campus,

PO: BCKV Campus Main Office, Mohanpur - 741252, Nadia, West Bengal, India

[2] School of Instrumentation, Devi Ahilya Vishwavidyalaya,
Khandwa Road, Indore-452017, India

[3]Magnetic and Superconducting Materials Section,
Raja Ramanna Centre for Advanced Technology, Indore-452013, India

## Abstract

Fe/Co bilayers were made using ion beam sputtering and then were subjected to nitrogen-ion irradiation to explore the influence of irradiation on their magnetic properties. Magnetization studies of Fe/Co bilayers along with their irradiated counterparts were then performed. The magnetization and coercivity of the as-deposited sample exhibited a two-fold uniaxial symmetry while the behavior for high temperature irradiated sample depicts a small shift in the two fold behavior. The magnetization data measured with the magnetic field perpendicular to the plane of the film exhibited a hysteresis in both the case of the irradiated sample as well as as-deposited sample. A small increase in coercivity for the irradiated sample was observed which is due to various reasons including formation of FeCo solid solution due to ion irradiation at the interface.

[*] e-mail: chiranjib@iiserkol.ac.in

## 1. Introduction

Exchange bias (EB) between ferromagnetic-antiferromagnetic (FM-AFM) bilayers, leading to shift and broadening of hysteresis loops when cooled below Neel temperature of AFM material is a well established phenomenon and was first discovered in the year 1956 by Meiklejohn and Bean (MB) [1]. Owing to immense application in data storage industries it still remains an active area of research worldwide. The physics behind exchange bias has been studied extensively for the last two decades, yet a satisfactory theoretical understanding is still lacking. A variety of experiments over different FM-AFM bilayers give us a plausible picture of EB phenomenon. The uncompensated AFM spins at the interface of FM-AFM is believed to be contributing to EB through short range exchange coupling between FM-AFM bilayers [2]. Soft magnetic Fe and hard anisotropic Co form a suitable exchanged coupled FM-AFM bilayer system to study EB at the interface. This system is known to exhibit exchange bias in its magnetization curves [Ref]. Ion-irradiation or thermal annealing is a well known process to modify the structure at the interface and hence the underlying magnetic property [3]. A number of irradiation studies were carried out using different high energy ions like Xe and Ar to modify the magnetic property of thin films [3]. Irradiation by nitrogen ions has a great impact on the enhancement of magnetisation as was first reported by Kim *et. al.* in the year 1972 [4]. FeN has a high saturation magnetization which coupled with its soft magnetic property makes this material a good candidate for inductive heads. Superconducting Quantum Interference Device (SQUID) is a powerful tool to do magnetic studies. In case of systems where small change in moment is significant SQUID is preferred as compare to other magnetic measurement technique. However it is very important that remnance field issue must be properly taken cared which in our case is done by recycling the magnetic field before every field. In the present work we have investigated the magnetisation study of Si(sub)/Co/Fe57/Fe/Al thin films irradiated by 50KeV nitrogen ions at different temperatures [5]. High temperature irradiated sample has shown an increase in coercivity as compared to the pristine or as deposited sample. For many types of applications we need switching both in the parallel and perpendicular directions. Since both the samples has shown out of plane magnetization, the results of nitrogen ion effect on out of plane magnetization will help us to tune the magnetization along this direction which will have many technological implications.

**2. Experimental**

Bilayers of Si/Co/Fe57/Fe/Al were deposited by ion beam sputtering technique. Ar ion beam was used for the deposition purpose on the Si 100 substrate. A base pressure of about $10^{-6}$ Torr was achieved prior to deposition. The total thickness of the films were 72nm including 32nm of Co and 35 nm of Fe. A 5nm film of Fe57 was deposited at the interface of Fe and Co for Mossbauer studies with a 5nm thick capping layer of Al to protect the film from oxidizing. Irradiation of films was done at 300$^{o}$C with 50KeV nitrogen ions by low energy accelerator having RF ion source, with a magnetic field of about 100 Oe being applied during irradiation to align the magnetization direction along the film plane. Using SRIM software the energy of nitrogen ions was calculated so that the nitrogen ion should cover only the interface Co/Fe57/Fe and ion beam mixing at the film/substrate interface could be avoided. Magnetization studies have been done by using Quantum design's Magnetic Property Measurement System (MPMS) at 5K, 100K and 300K by varying the angle between the field direction and anisotropy axis. These were performed at an interval of approximately 45$^{o}$. We have cycled the magnetic field to get rid of any trapped field in the superconducting magnet before carrying out the different hysteresis loops. All the results are repeatable.

**3. Results and Discussions**

Both the pristine and high temperature irradiated samples have shown positive exchange bias when the magnetisation was measured as the function of the angle that the anisotropy axis makes with the external field. This was achieved by carefully rotating the sample at an interval of 45 degree with respect to field direction (in-plane) and measuring data at various temperatures down to 5K from room temperature. In order to maintain accuracy this interval was chosen which in lesser interval case will be difficult to maintain. The samples has shown EB and broadening of hysteresis loops at low temperature owing to an increase in coercivity brought about by the reduced thermal excitation of the trapped domain wall. Variation of coercivity was observed when samples were rotated in-plane with respect to the magnetic field indicating that the anisotropy axis lies in the plane of sample. The highest coercivity of about 22 Oe for high temperature implanted sample was observed at around 135$^{o}$. During deposition the bottom Co layer gets aligned in plane because of the applied magnetic field, the Fe layer deposited subsequently gets aligned to the bottom layer because of strong coupling with Co layer having high anisotropic energy density. Thus the system is expected to possess low anisotropy dispersion which causes easy domain wall motion and hence low coercivity was observed in the as deposited sample [6]. Single phase magnetic behaviour as shown in fig 1(a) was observed because of strong ferromagnetic exchange coupling between the layers and film thickness is of the order of domain wall thickness. Coercivity being an intrinsic magnetic property is highly affected by defects caused by ion irradiation [7]. A small increase in coercivity as shown in fig 2 at the high temperature implanted sample was observed owing to formation of FeCo solid solution at the interface [8] as certified by higher value of hyperfine field in the Mossbauer data of same sample (not shown here) which clearly shows the diffusion of Co into Fe lattice at the interface. Ion irradiation is a well reported technique for creating defects and hence modification of magnetic properties. Being very sensitive to defects created by ion irradiation, coercivity was also found to increase because of this reason. This also verifies that stress in the film caused by irradiation is not relaxed fully even at high temperature. Magnetisation curves taken with the field lying in the plane perpendicular to film surface sample has shown hysteresis loop with coercivity of 180 Oe and 194 Oe for both pristine and irradiated sample measured at 300K as in fig 1(b). This suggests out of plane anisotropy for both the samples which also shows a small concomitant shift in the in-plane anisotropy as is evident from fig 4. Positive exchange bias shown by both the samples indicated the presence of AF coupling at the Fe/Co interface. Magnetic measurement revealed that the magnetization and coercivity posses clear two-fold symmetric for the pristine sample while the high temperature irradiated sample has shown deviation from two fold symmetry with its axis appears to have rotated. In over previous studies over the same systems we have not find any change in two fold anisotropy axis when the system was irradiated at different nitrogen fluence [9]. This might be because SQUID is usually very sensitive to even very small change in moment and is difficult to detect in Magneto Optic

Kerr Effect (MOKE) measurement which is a surface effect. This is more evident from the fact that sharp switching behaviour which was observed using SQUID for the perfectly (0, 90, 180 and 270 angle positions) in-plane aligned (with respect to field) as-deposited sample is observed in the irradiated sample placed at an angle integral of 45 degree (fig. 1a inset) which is clear indication of symmetry axis rotation facilitated by ion irradiation. The MOKE measurement of the irradiated sample measured at an interval of 10 degree did not revealed any uniaxial symmetry. However positive exchange bias is also experienced in MOKE measurement for this sample. While comparing the two measurements we found that for particular angle say 90 degree we saw a marked change in hysteresis loop having round edges indicating continues spin aligning with magnetic field which was not seen at any angle in MOKE measurement showing squared loops. This might be because of high resolution of SQUID magnetometer capable of measuring moment at a very small field increment. Shift in the coercivity axis indicate some structural change caused by ion irradiation at the interface. Because of $N_2$ ion irradiation some structural change of hcp Co to fcc Co has been observed for the same system irradiated at 300$^o$C [10]. Earlier Zhang *et al.* had observed a structural transformation of hcp Co into fcc Co because of Xe ion irradiation of Substrate/Fe/Co system. However no such change was observed in similar geometry of film while a transformation occurs after in-situ annealing at 300$^o$C on Substrate/Co/Fe [3].

**4. Conclusion**

To conclude, we demonstrated the effect of high temperature $N_2$ ion irradiation over Co/Fe (Fe on top) bilayers in inducing a shift in anisotropy axis. Single phase magnetic behaviour was observed in the films arising from strong exchange interaction between the ferromagnetic layers. A small increase in coercivity was observed in the high temperature irradiated sample which indicates the formation of FeCo alloy at the interface because of intermixing. Magnetisation data reveals that as deposited sample posses two fold anisotropy and after irradiation the two fold symmetry found to be shifted by a small angle. Also for pristine sample the coercivity was two-fold symmetric while after irradiation it was found to be rotated by small angle indicating some structural change induced by irradiation. The out of plane magnetization for the high temperature irradiated sample certainly opens avenues for better technological applications by making the out of plane magnetization available for manipulation and data storage.

The authors acknowledge the help of Mr. Poddar for deposition of films.

References:

[1] W. H. Meiklejohn and C. P. Bean, Phys. Rev. **102**, 1413 (1957).

[2] J. Nogués and I. K. Schuller, J. Magn. Magn. Mater. **192**, 203, (1999).

[3] K. Zhang, R. Gupta, G. A. Muller, P. Schaaf, and K. P. Lieb, Appl. Phys. Lett. **84**, 3915 (2004).

[4] T.K. Kim and M. Takahashi, Appl. Phys. Lett. **20**, 492, (1972).

[5] The reason for using Fe57 was to do some Mossbauer studies which will be reported elsewhere.

[6] M.H. Park, Y.K. Hong, S.H. Gee, M.L. Mottern and T.W. Jang, J. Appl. Phys., **91**, 7218, (2002).

[7] Ratnesh Gupta, K.H. Y Han, G.A. Müller, P. Schaaf, K. Zhang, K.P. Lieb, J. Appl. Phys. **97**, 073911, (2005).

[8] Deeder Aurongzeb, K. Bhargava Ram, and Latika Menon, Appl. Phy. Lett., **87**, 172509 (2005).

[9] Ratnesh Gupta, Ashish Khandelwal, Raisa Ansari, Ajay Gupta, K.G.M. Nayer, Surface & Coatings Technology, **203**, 2713-1720, (2009).

[10] Ratnesh Gupta, Ashish Khandelwal, Raise Ansari, K.G.M Nair, W. Leitenberger, U. Pietsch, D.M. Phase, Ajay Gupta, Nucl. Instrum. And Methods B **266**, 1705, (2008).

Figures:

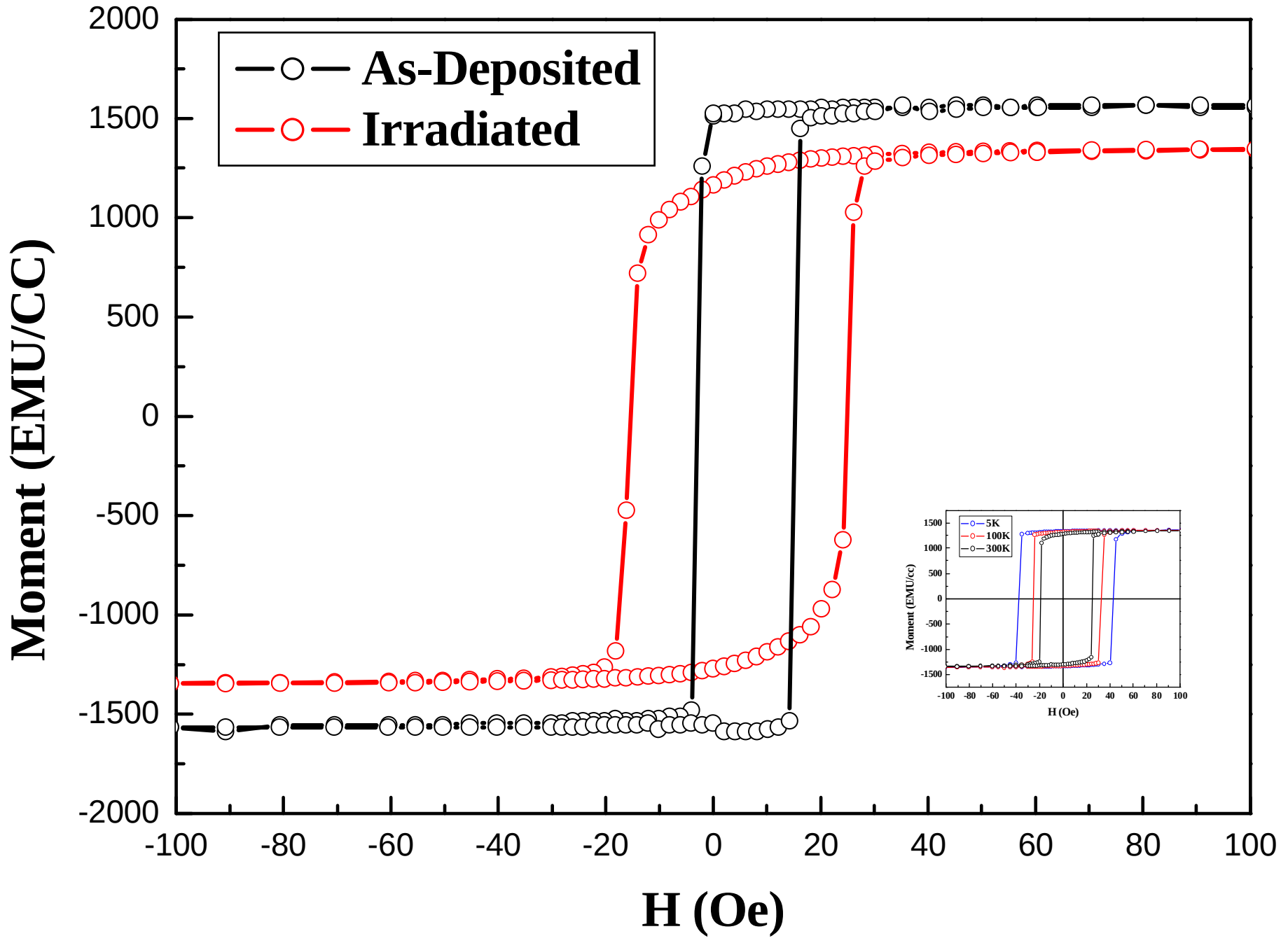

Fig 1(a) shows the in-plane hysteresis loop for the as-deposited (-□-) and irradiated (-○-) sample measured at 300K indicating the single phase switching.

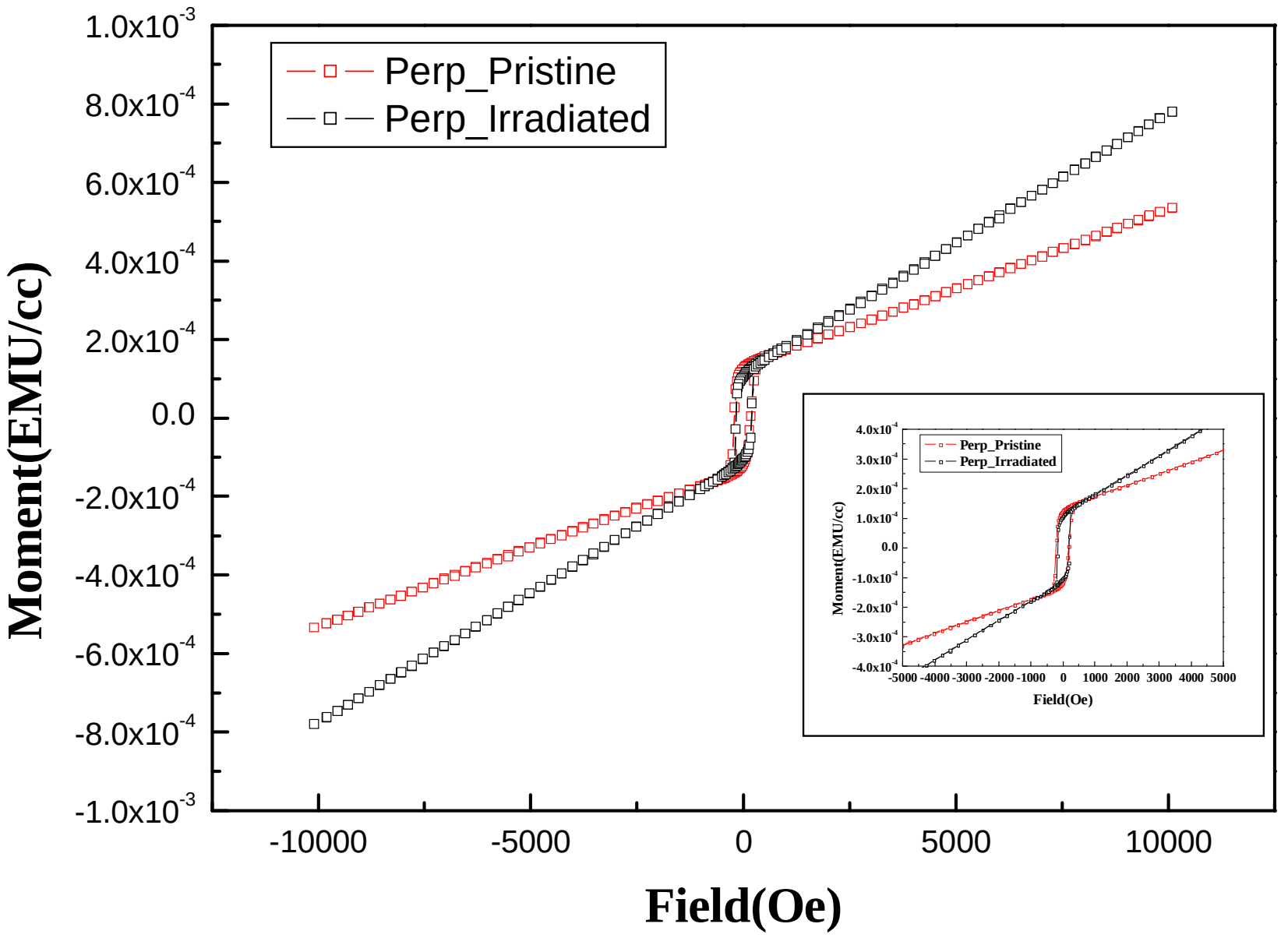

Fig 1(b) shows the out of plan hysteresis loops for the as-deposited (-□-) and irradiated sample (-○-). The irradiated sample is showing hysteresis while no well defined hysteresis is exhibited by as-deposited sample.

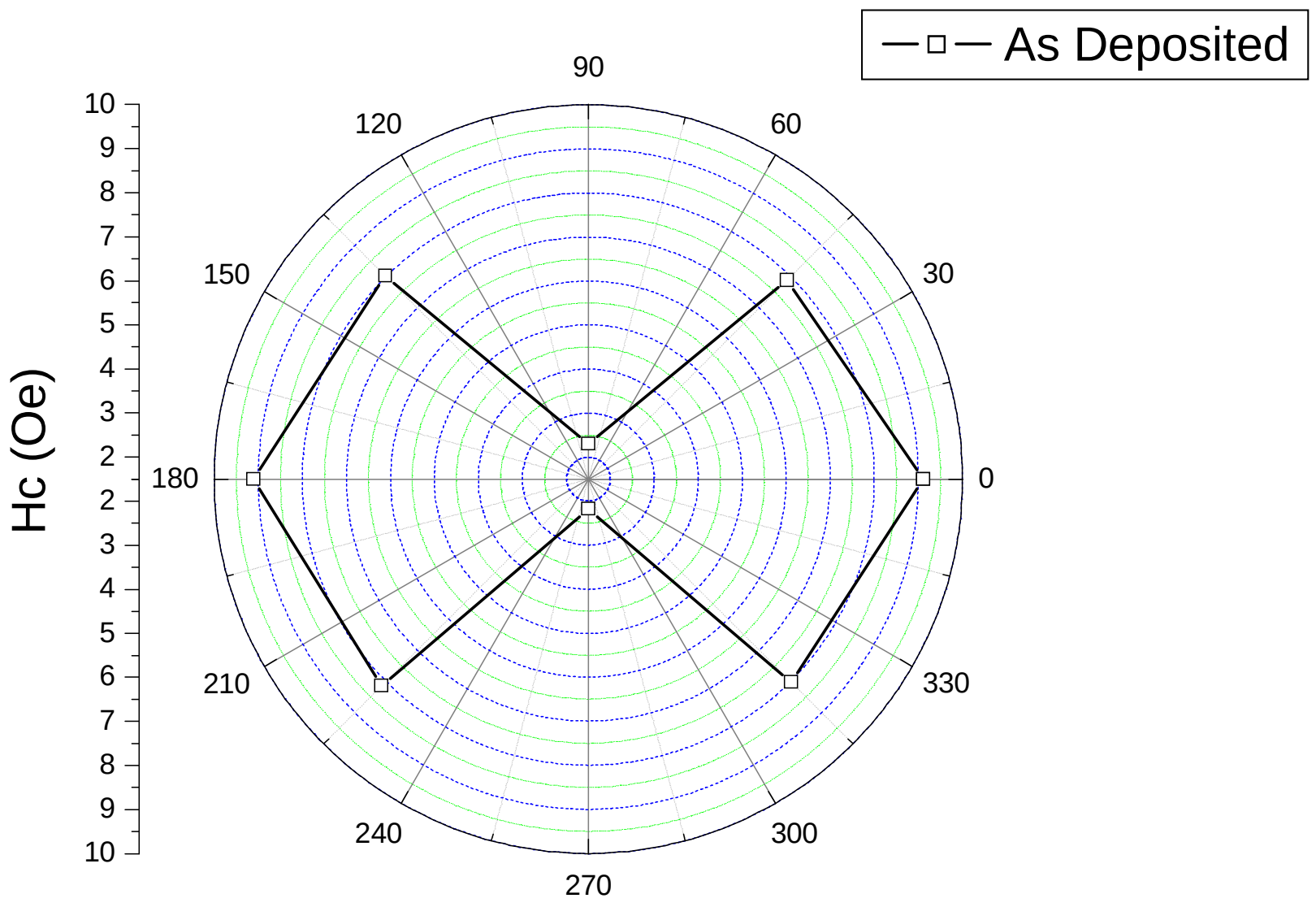

Fig 3 Polar diagram of coercivity of as-deposited sample showing two-fold symmetry.

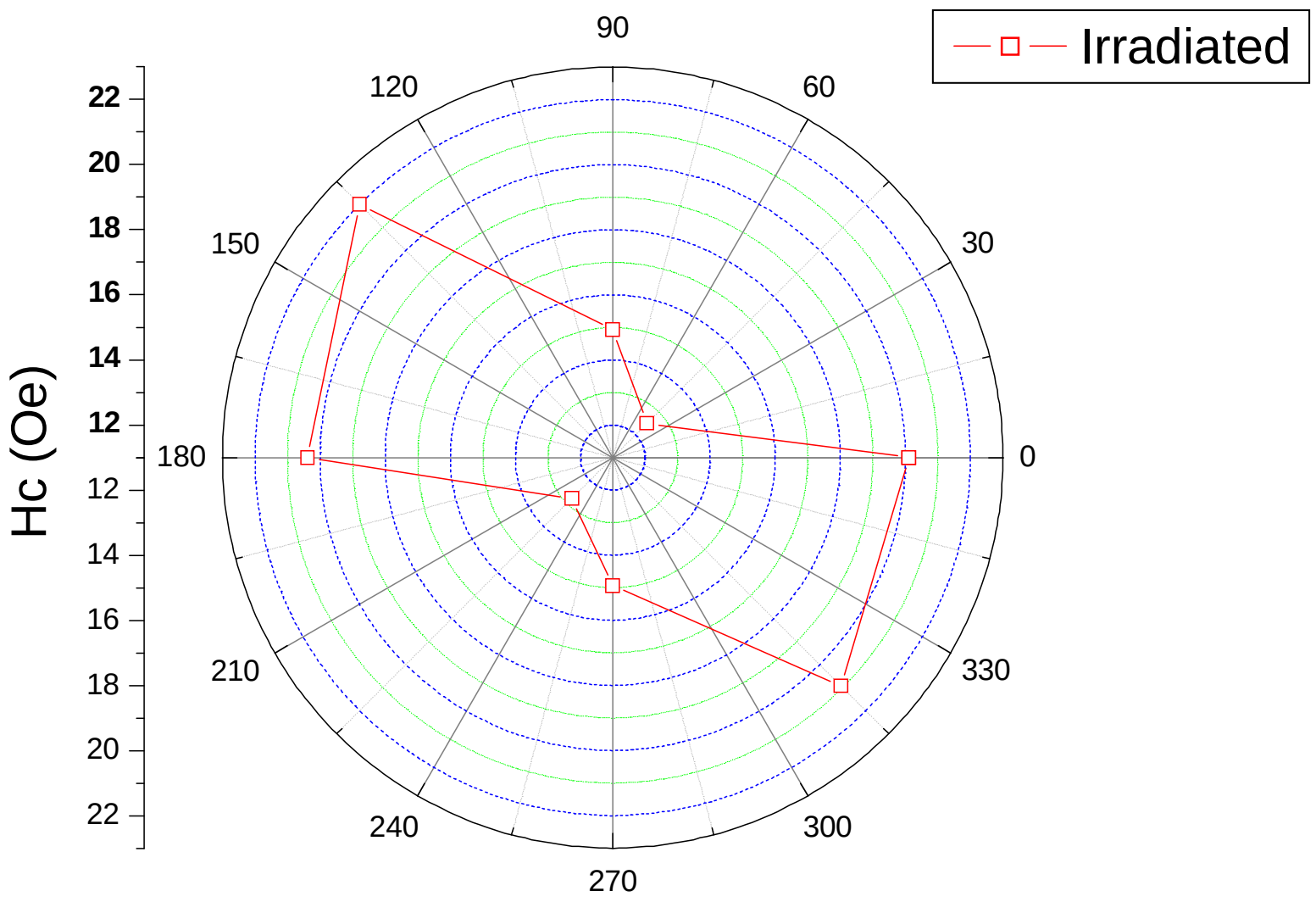

Fig 4. Polar diagram of irradiated sample indicating the tilted two fold symmetry

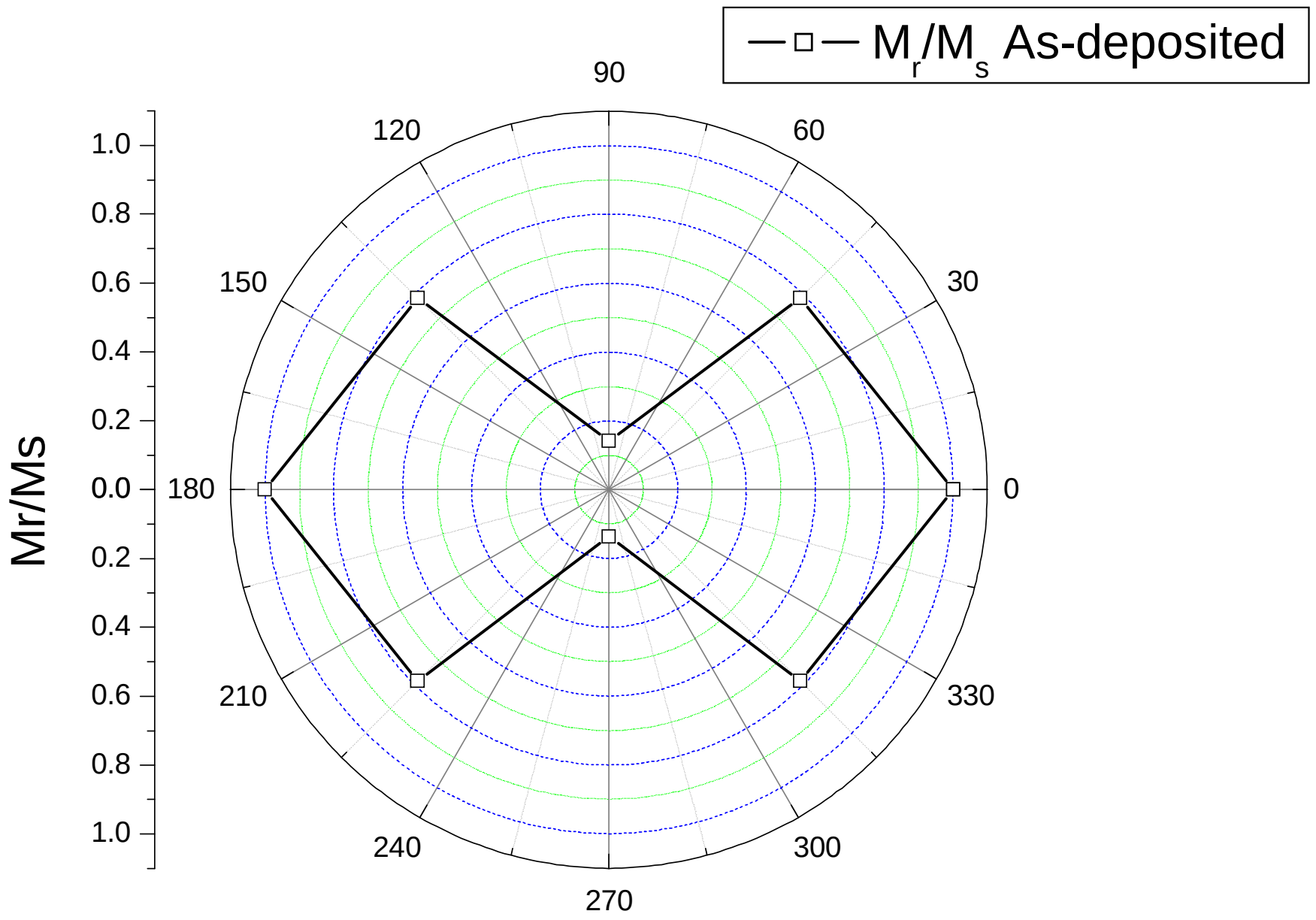

Fig 5. Polar diagram of the remnant magnetization ratio Mr/Ms of as-deposited sample showing two fold symmetry.

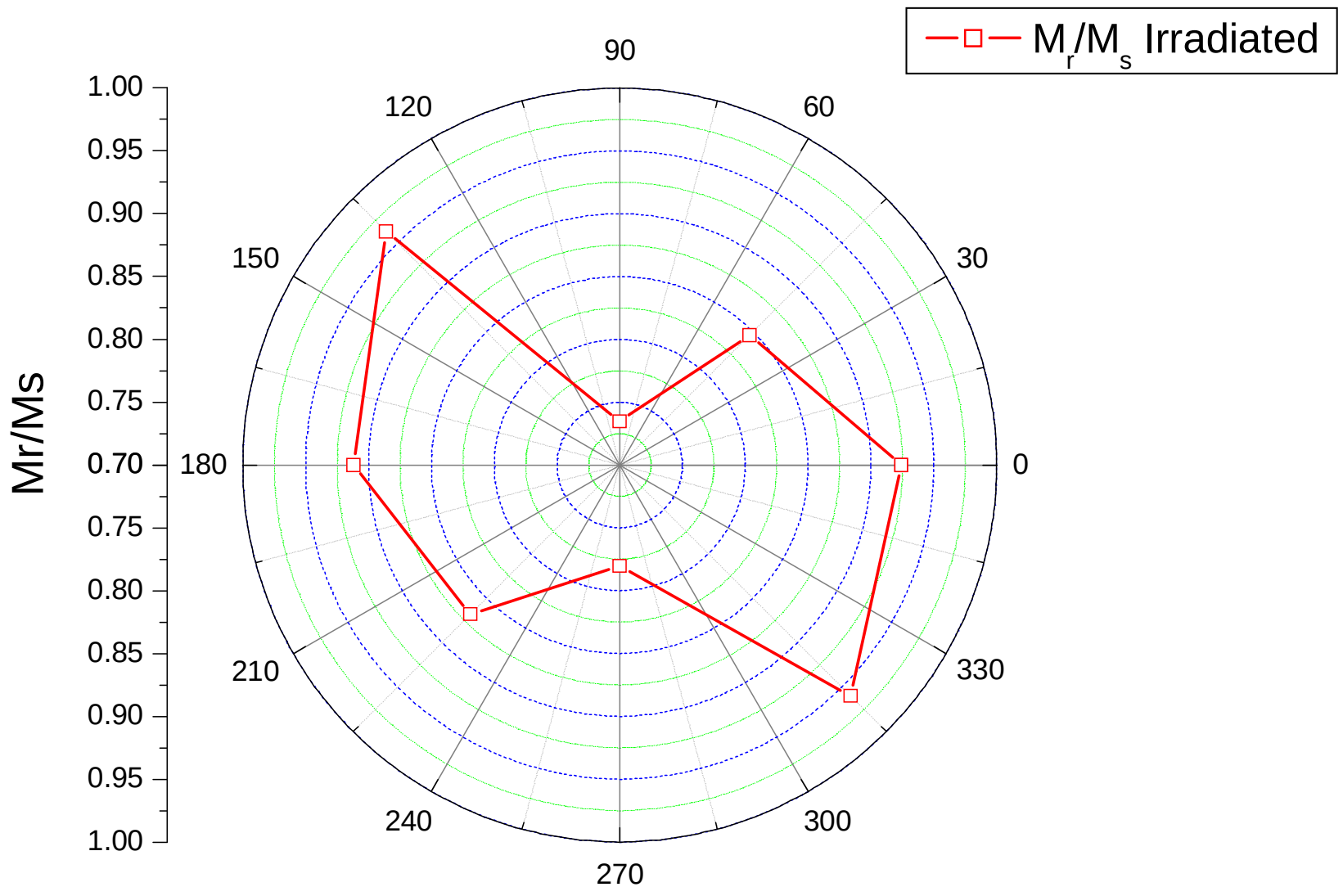

Fig 6. Polar diagram of the remnant magnetization ratio Mr/Ms of irradiated sample showing a small change in the two fold symmetry.